\documentclass[12pt]{article}
\usepackage[a4paper, margin=1in]{geometry}

\usepackage[english]{babel}
\usepackage{amsmath, amsfonts, amssymb, dsfont, graphicx, tabularx, adjustbox, graphics, bbm, mathrsfs, mathtools, multirow, nameref, natbib, setspace, thmtools, bm, algorithm, algpseudocode}
\usepackage{enumitem}
\usepackage[font=footnotesize,justification=centering]{caption}
\usepackage{subcaption}

\usepackage[normalem]{ulem}

\usepackage{hyperref, tabularx, booktabs}

\DeclareMathOperator*{\argmin}{argmin}

\usepackage{authblk}

\begin{document}
\def\spacingset#1{\renewcommand{\baselinestretch}%
{#1}\small\normalsize} \spacingset{1}

\title{Shared Hidden‑factor Information Framework for Multiple Behavioral Tasks}

\author{Yuan Bian$^{1}$, Yuanjia Wang$^{1,2}$, and Xingche Guo$^{3,4}$\\
$^{1}$Department of Biostatistics, Columbia University\\
$^{2}$Department of Psychiatry, Columbia University\\
$^{3}$Department of Statistics, University of Connecticut\\
$^{4}$Institute for the Brain and Cognitive Sciences, University of Connecticut\\
}
\date{}

\maketitle

\begin{abstract}
Understanding cognitive processes in major depressive disorder (MDD) often relies on behavioral tasks, which are typically analyzed separately, overlooking potential correlations and shared latent structure. To address this limitation, we propose the \textbf{S}hared \textbf{H}idden‑factor \textbf{I}nformation \textbf{F}ramework for Multiple Behavioral \textbf{T}asks (SHIFT), a joint modeling approach that leverages shared information across tasks, allowing each task to benefit from information learned by the others. SHIFT introduces subject-specific latent factors that capture cross-task dependencies while accommodating individual heterogeneity in decision-making, response times (RTs), and strategy switching. To address computational challenges without requiring high-dimensional integration, we develop an expectation–maximization with variational approximation algorithm that preserves both temporal structure and between-task dependencies. Through extensive simulation studies, we demonstrate that SHIFT substantially improves estimation accuracy and efficiency relative to single-task analyses. We then apply SHIFT to a study of MDD to jointly model the Probabilistic Reward Task (PRT) and the Flanker Task (FT). Results indicate that MDD participants show lower engagement in the PRT and reduced focus in the FT compared with healthy controls. Moreover, when individuals are engaged and focused, they exhibit longer RTs. Although observed RTs do not predict treatment response, the shared parameters recovered by SHIFT showed suggestive treatment‐modulation patterns, indicating their potential as exploratory behavioral markers for therapeutic outcomes.\\
\end{abstract}

\noindent%

\doublespacing
\allowdisplaybreaks

\section{Introduction}
Depressive disorder affects approximately $4$\% of the global population, corresponding to about $332$ million people, and is a major contributor to disability and mortality \citep{whiteford2013global,gbd2021}. Despite substantial clinical efforts, effective treatment strategies remain limited \citep{trivedi2016establishing}, partly due to patient heterogeneity that is not captured by traditional symptom-based diagnoses. To address this limitation, the National Institute of Mental Health’s Research Domain Criteria (RDoC) framework \citep{insel2010research} advocates redefining mental disorders using latent constructs derived from biological and behavioral measures rather than relying solely on clinical symptoms. Within this framework, behavioral tasks combined with computational methods \citep{huys2016computational} can identify markers of underlying cognitive and neural mechanisms and improve characterization of mental disorders beyond traditional approaches \citep{passos2019big}. However, these approaches also introduce statistical challenges in modeling complex multivariate behavioral data and individual heterogeneity, motivating the development of more flexible computational frameworks.

\begin{figure}[h!]
    \begin{subfigure}{0.5\textwidth}
         \centering
         \includegraphics[width=\textwidth]{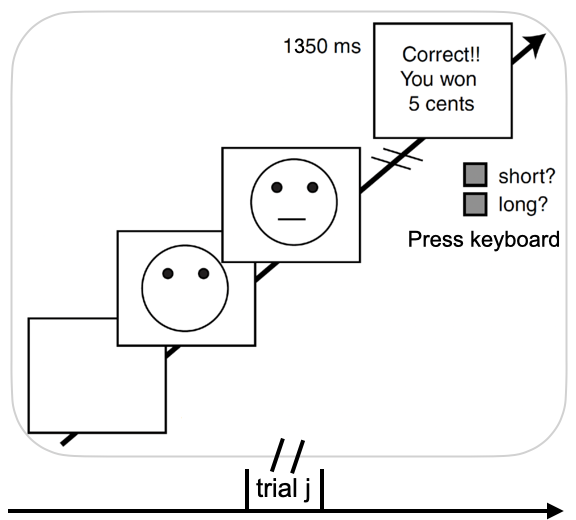}
         \caption{The Probabilistic Reward Task (PRT)}
         \label{Fig1a}
    \end{subfigure}
    \begin{subfigure}{0.5\textwidth}
         \centering
         \includegraphics[width=\textwidth]{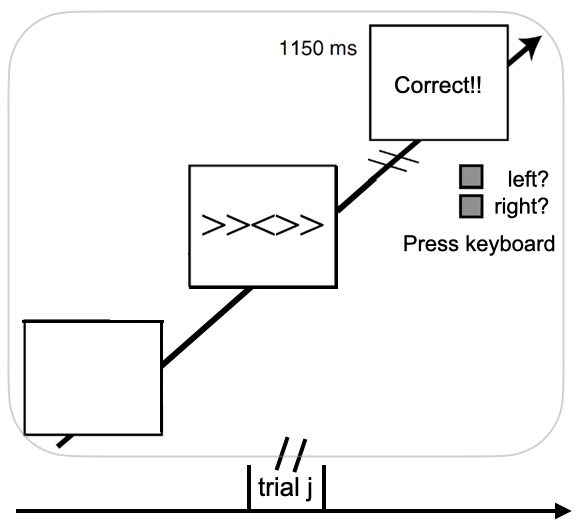}
         \caption{The Flanker Task (FT)}
         \label{Fig1b}
    \end{subfigure}
    \begin{subfigure}{\textwidth}
    \centering
    \vspace{5mm}
    \includegraphics[width=0.75\linewidth]{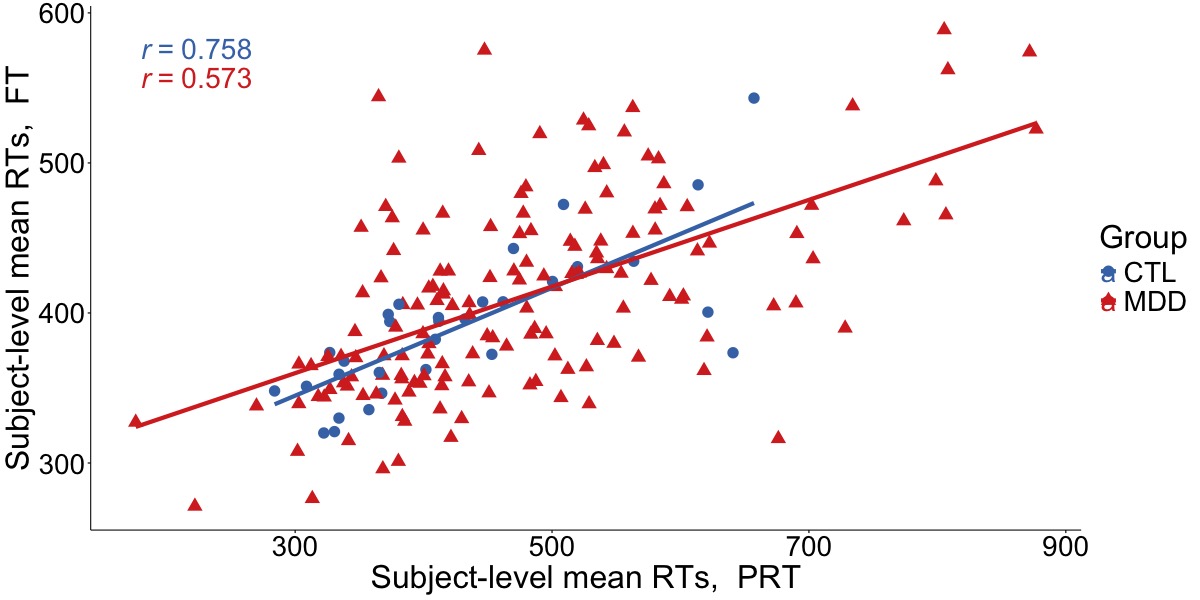}
    \caption{Scatter plot of response times (RTs) for controls (CTL) and major depressive disorder (MDD) across the PRT and FT.}
    \label{Fig1c}
    \end{subfigure} 
    \caption{PRT and FT schematics and RT comparison across groups.}
\end{figure}

The \emph{Establishing Moderators and Biosignatures of Antidepressant Response for Clinical Care} (EMBARC) study \citep{trivedi2016establishing} is a landmark study of MDD that integrates multiple data modalities and a suite of behavioral tasks to probe cognitive and affective processes in individuals with MDD and healthy controls (CTLs). We focus on the behavioral tasks, starting with the \emph{Probabilistic Reward Task} \citep[PRT;][]{pizzagalli2005toward}, which measures a subject's ability to modify behavior based on reward contingencies.
Each PRT session comprised $200$ trials, administered in two blocks of $100$ trials separated by a $30$-second break. 
On each trial, participants identify a briefly presented cartoon face as having a short or long mouth by pressing one of two buttons. To induce a response bias, correct identifications of the short mouth (the ``rich'' stimulus) are rewarded more frequently than correct identifications of the long mouth (the ``lean'' stimulus). Participants are instructed to maximize their total rewards, with a subtle difference in mouth length adding a perceptual challenge that often biases responses toward the more frequently rewarded stimulus. See Figure~\ref{Fig1a} for a schematic. Reinforcement learning \citep[RL;][]{sutton2018reinforcement} and drift-diffusion models \citep[DDMs;][]{ratcliff2008diffusion} have been used to characterize behavioral adaptation and decision dynamics in the PRT \citep{bian2025ddm}.

A second task we model in this work is the \emph{Flanker Task} (FT), a canonical conflict task for quantifying selective attention, inhibitory control, and executive function \citep{holmes2010serotonin}. Engagement-related features extracted from the PRT were found to be associated with behavioral metrics from the FT \citep{bian2025ddm}. On each trial, participants are required to indicate the direction of a central target arrow while disregarding the surrounding flanker arrows. Trials are labeled ``congruent'' when the target and flankers point in the same direction (e.g., ``$>>>>>$'') and ``incongruent'' when they point in competing directions (e.g., ``$>><>>$''). A typical trial comprises fixation, stimulus onset, and a response window (see Figure~\ref{Fig1b}), and participants are instructed to respond quickly and accurately to the target. The FT involved $5$ blocks of $70$ trials, with $46$ congruent and $24$ incongruent trials for each block.
A robust congruency effect is typically observed for conflict tasks: response times (RTs; i.e., the latency between stimulus onset to action) are faster and error rates are lower on congruent than on incongruent trials. \cite{dillon2015computational} and \cite{ulrich2015automatic} used multiple DDMs to model conflict tasks.

Existing approaches for analyzing behavioral tasks are restricted to analyzing a single task type at a time. Yet when the same individuals complete multiple tasks, modeling each separately ignores cross–task correlations and shared cognitive mechanisms \citep{caruana1997multitask}. As shown in Figure~\ref{Fig1c}, our preliminary analyses reveal a positive association between individuals' RTs in the PRT and FT. Specifically, the Pearson correlation is $0.573$ for the MDD group, with 95\% confidence interval $(0.456, 0.671)$, and $0.758$ for the CTL group, with 95\% confidence interval $(0.552, 0.877)$; both $p$-values are less than 0.001. These estimates indicate a moderate-to-strong positive relationship in MDD and a strong positive relationship in CTL, suggesting shared underlying processes across tasks. 
Joint modeling of multiple tasks with shared structures, commonly referred to as multi-task learning \citep[MTL;][]{caruana1997multitask}, has become increasingly popular in modern applications \citep{zhang2018overview}.
Our goal is therefore to develop a joint modeling framework that identifies latent factors driving behaviors across tasks while borrowing strength through their shared structure.

In this paper, we propose a generalized factor model framework, the \emph{\textbf{S}hared \textbf{H}idden‑factor \textbf{I}nformation \textbf{F}ramework for Multiple Behavioral \textbf{T}asks} (SHIFT), which enables knowledge transfer across tasks, allowing each task to benefit from information learned by the others. Our framework builds on two established paradigms. First, factor analysis, a widely used method in the social and behavioral sciences, employs a small number of latent factors to capture variation across a large number of observed variables \citep{bartholomew2011latent}. Second, hard parameter sharing \citep{ruder2017overview}, the most common approach to MTL in neural networks \citep{caruana1993multitask}, assumes that some parameters are shared across tasks while others remain task-specific. SHIFT extends both frameworks to multiple behavioral tasks, capturing individual variations in decision-making, RTs, and dynamic strategy switching.

Estimation for the SHIFT is challenging due to two sources of latent structure: (i) trial-dependent latent states capturing decision dynamics, and (ii) subject-specific factors that enable information sharing across tasks. Joint estimation requires integrating over high-dimensional, temporally dependent variables, making both direct maximum likelihood and classical expectation–maximization \citep[EM;][]{dempster1977maximum} algorithms infeasible. Variational approximation \citep[VA;][]{blei2017variational} offers a tractable alternative, but standard mean-field assumptions break the temporal dependence and introduce bias. To address this, we develop an EM–VA algorithm with structured partial mean-field approximation, leveraging forward–backward recursions and limiting variational parameters. While the EM–VA connection is well established \citep{neal1998view, jordan1999introduction} and used in other contexts \citep[e.g.,][]{zeng2023zero}, our contribution is to adapt and extend it to models with coupled temporal and subject-level latent structure for behavioral tasks.

Our approach also differs from existing approaches designed for related methods. For example, mixed hidden Markov models \citep{altman2007mixed} incorporate similar structures but rely on computationally intensive Monte Carlo EM and are typically restricted to exponential family models. In contrast, our framework accommodates the Wiener first-passage time distribution and enables efficient estimation via variational methods. \cite{duan2023adaptive} proposed adaptive MTL methods, a form of soft parameter sharing \citep{ruder2017overview}, that account for task similarity by penalizing parameter differences. However, such methods often introduce additional computational burden through hyperparameter tuning and may require data splitting for training, validation, and testing. Structured VA \citep{hoffman2015structured} allows local variational factors to depend on global parameters, while hierarchical VA \citep{ranganath2016hierarchical} enriches dependence through flexible variational priors (e.g., mixtures or normalizing flows). Although these approaches improve flexibility, they do not directly address the temporal dependence inherent in sequential latent structures. By contrast, the SHIFT EM–VA estimator explicitly preserves sequential structure while sharing information across tasks, providing a principled and efficient solution.

The remainder of this paper is organized as follows. Section~\ref{sec: m} introduces the proposed framework for joint modeling of multiple behavioral tasks, while Section~\ref{sec: ep} details the parameter estimation algorithm. Section~\ref{sec: ss} presents simulation studies to evaluate the performance of the proposed method. In Section~\ref{sec: aes}, we demonstrate the application to the PRT and FT in the EMBARC study. We discuss potential extensions for future research in Section~\ref{sec: d}.

\section{Method}\label{sec: m}
Consider $N$ individuals participating in $K$ behavioral tasks, where participants repeatedly select actions in response to the presented stimuli under time constraints, sometimes receiving feedback or rewards. Each individual $i\in\{1,\dots,N\}$ is characterized by $p$-dimensional baseline covariates $\boldsymbol{X}_i\in\mathbb{R}^p$. For task $k\in\{1,\dots,K\}$, let $\boldsymbol{\mathcal{D}}_i^{(k)} = ( \mathcal{D}_{i1}^{(k)}, \dots,  \mathcal{D}_{iJ_{ik}}^{(k)})$ denote the sequential data for individual $i$, comprising stimulus $S_{ij}^{(k)}$, actions $A_{ij}^{(k)}$, and RTs $T_{ij}^{(k)}$ over trials $j=1,\dots,J_{ik}$. For reward-learning paradigms, $\mathcal{D}_{ij}^{(k)}$ additionally includes rewards $R_{ij}^{(k)}$. In this paper, only the participants' outcomes (i.e., actions and RTs) are treated as random variables, while task stimuli (and rewards) are regarded as fixed exogenous conditioning variables and are not modeled probabilistically.

For the PRT, we have $\mathcal{D}_{ij}^{(1)} = \big(S_{ij}^{(1)}, \ A_{ij}^{(1)}, \ T_{ij}^{(1)}, \ R_{ij}^{(1)}\big)$,
where $S_{ij}^{(1)}\in\{0,1\}$ indicates a lean ($0$) or rich ($1$) reward stimuli, $A_{ij}^{(1)}\in\{0,1\}$ denotes selection of the lean ($0$) or rich ($1$) option, and $R_{ij}^{(1)}\in\{0,1\}$ records trial-level reward. For the FT, $\mathcal{D}_{ij}^{(2)} = \big(S_{ij}^{(2)}, \ A_{ij}^{(2)}, \ T_{ij}^{(2)}\big)$,
where $S_{ij}^{(2)}\in\{0,1\}$ denotes an incongruent ($0$) or congruent ($1$) stimulus, and $A_{ij}^{(2)}\in\{0,1\}$ indicates an incorrect ($0$) or correct ($1$) response.
While we focus on these two types of behavioral tasks in this paper, the proposed method applies to a broader range of task paradigms. 

\subsection{Modeling decision-making dynamics within a single task}\label{subsec: mdmdst}
Within each task $k$, the drift-diffusion models \citep[DDMs;][]{ratcliff2008diffusion} are commonly used to jointly model actions $A_{ij}^{(k)}$ and RTs $T_{ij}^{(k)}$. DDMs posit that choices arise from the noisy accumulation of evidence over time. Evidence starts at $X(0)=\beta\alpha$ with $0<\beta<1$, between a lower bound $0$ and an upper bound $\alpha>0$; a decision is made when the process first hits either bound. The latent decision variable $X(t)$ follows a drift–diffusion process with constant drift $v\in\mathbb R$ and diffusion $\sigma$: $dX(t)=v\,dt+\sigma\,dB(t)$, 
where $B(t)$ is a standard Brownian motion. For identifiability, it is common to fix $\sigma=1$.
To account for processes unrelated to the decision itself, such as stimulus perception and motor execution, a non-decision time $\tau>0$, is added to the total RT.

A growing literature shows that individuals may switch between distinct strategies across trials \citep[e.g.,][]{ashwood2022mice, li2024dynamic, bian2025ddm, guo2025hmm}. To capture these dynamics, we introduce a latent strategy variable $U_{ij}^{(k)}\in\{0,1\}$ indicating the decision-making state (e.g., engaged versus lapsed) used by subject $i$ at trial $j$ in task $k$.
Conditional on the strategy mode $U_{ij}^{(k)}$, the current stimuli $S_{ij}^{(k)}$, and history $\mathcal{H}_{ij}^{(k)}=\{S_{il}^{(k)},A_{il}^{(k)},T_{il}^{(k)},R_{il}^{(k)}\}_{l=1}^{j-1}$, the joint distribution of the RT and decision follows a mixture of DDMs:
\begin{align}
p\big(T_{ij}^{(k)},A_{ij}^{(k)}\mid U_{ij}^{(k)}, S_{ij}^{(k)}, \mathcal{H}_{ij}^{(k)}\big)=&\ I\big(U_{ij}^{(k)}=0\big)p\big(T_{ij}^{(k)},A_{ij}^{(k)};\alpha_{ik0},\beta_{ik0},v_{ik0j},\tau_{ik0}\big)\notag\\
&+I\big(U_{ij}^{(k)}=1\big)p\big(T_{ij}^{(k)},A_{ij}^{(k)};\alpha_{ik1},\beta_{ik1},v_{ik1j},\tau_{ik1}\big),\label{eq: fta}
\end{align}
where $p(t,a; \alpha, \beta, v, \tau)$ is the joint density function for the Wiener first-passage time distribution \citep{ratcliff1978diffusion}, with details provided in Web Appendix A. The drift $v_{ikuj}$ may depend on $(S_{ij}^{(k)},\mathcal{H}_{ij}^{(k)})$ via a parametric or learned function, allowing task- and history-dependent evidence accumulation.
The latent engagement state $\{ U_{ij}^{(k)} \}_{j=1}^{J_{ik}}$ evolves according to a first-order hidden Markov model (HMM) with covariate-dependent transitions:
\begin{equation}
\label{eq: hmm}
\operatorname{logit}\Pr\big(U_{i,j+1}^{(k)}=1\mid U_{ij}^{(k)}=l\big)=\gamma_{kl0}+\boldsymbol\gamma_{kl1}^\top \boldsymbol X_i \ , \quad l=0, 1,
\end{equation}
and initial distribution $\Pr(U_{i1}^{(k)}=l)=\pi_{kl}$ with $\pi_{k1} \in (0, 1)$ and $\pi_{k0} = 1 - \pi_{k1}$. 

In the EMBARC application, for the PRT ($k=1$), we interpret $U_{ij}^{(1)}=1$ as an ``engaged'' mode in which the participant deliberately accumulates evidence in accordance with a reinforcement-learning (RL) scheme, and $U_{ij}^{(1)}=0$ as a ``lapsed'' mode reflecting inattentive responding, characterized by random guessing.
Under the engaged mode ($U_{ij}^{(1)}=1$), the drift rate is linked to trial-wise value differences: $v_{i11j}
= c \Big\{ Q_{ij}\big(1,S_{ij}^{(1)}\big) - Q_{ij}\big(0,S_{ij}^{(1)}\big) \Big\}$, where $c>0$ is a scaling (reward-sensitivity) parameter and $Q_{ij}(a,s)\;=\;\mathbb{E} \big( R_{ij}^{(1)} \mid A_{ij}^{(1)}=a,\ S_{ij}^{(1)}=s \big)$
denotes the expected reward for action $a$ in stimuli $s$. We update values via a standard Q-learning rule for the realized pair $(a_{ij}^{(1)},s_{ij}^{(1)})$: $Q_{i,j+1}(a,s)
= Q_{ij}(a,s) + b\,\Big\{ R_{ij}^{(1)} - Q_{ij}(a,s) \Big\}\,\mathbbm{1}\!\left\{(a,s)=(a_{ij}^{(1)},s_{ij}^{(1)})\right\}$,
with learning rate $b\in(0,1)$.
Under the lapsed mode ($U_{ij}^{(1)}=0$), we fix $v_{i10j}=0$ and $\beta_{i10}=0.5$, yielding a choice probability $\Pr\big(A_{ij}^{(1)}=1 \mid U_{ij}^{(1)}=0\big)=0.5$ consistent with random guessing \citep{bian2025ddm}.

For the FT ($k=2$), we similarly interpret $U_{ij}^{(2)}=1$ as a ``focused'' mode and $U_{ij}^{(2)}=0$ as a ``reduced'' (low-engagement) mode, in which participants are more susceptible to distraction.
In the focused mode ($U_{ij}^{(2)}=1$), the drift rate is decomposed into controlled and automatic components: $v_{i21j}
= v_c \;+\; \operatorname{sgn}(S_{ij}^{(2)}-\tfrac{1}{2}) \ v_a$,
where $v_c>0$ reflects controlled, goal-directed processing and $v_a>0$ captures automatic processing. The term $\operatorname{sgn}(S_{ij}^{(2)}-\tfrac{1}{2})$ flips the sign across trial types, so that on incongruent trials ($S_{ij}^{(2)}=0$) the automatic component opposes controlled evidence (reducing the drift), whereas on congruent trials ($S_{ij}^{(2)}=1$) the two components align (increasing the drift).
In the reduced mode ($U_{ij}^{(2)}=0$), automatic processing persists while controlled processing is attenuated: $v_{i20j}
= \rho\, v_c \;+\; \operatorname{sgn}(S_{ij}^{(2)}-\tfrac{1}{2}) \ v_a$ with $\rho \in (0,1)$.
We also take $\beta_{i20}=\beta_{i21}$, reflecting that the initial bias is unlikely to differ across engagement states. This specification is consistent with empirical findings that automatic responses persist under lapses of attention, whereas goal-directed control weakens, producing stronger conflict effects under distraction or fatigue \citep{lorist2012trial}. 

We refer to these methods as the \textbf{S}eparate \textbf{P}arameter \textbf{L}earning \textbf{I}n behavioral \textbf{T}asks (SPLIT). The SPLIT for the PRT was introduced in \cite{bian2025ddm}, while its extension to the FT is new in this work.

\subsection{Multiple behavioral tasks modeling framework}\label{subsec: mbtmf}
To jointly model $K$ behavioral tasks, facilitate knowledge transfer across tasks, and allow each task to benefit from information learned in others, we propose a generalized factor model framework. The key idea is to couple task-specific parameters through subject-level latent factors $\boldsymbol{f}_i$ that influence multiple tasks simultaneously. We refer to this approach as the \textbf{S}hared \textbf{H}idden-factor \textbf{I}nformation \textbf{F}ramework for multiple behavioral \textbf{T}asks (\textsc{SHIFT}).

Let $\boldsymbol \theta_{ik}\in\mathbb{R}^{d_k}$ denote the vector of subject– and task-specific parameters encoding individual cognitive characteristics in task $k$ for subject $i$ (e.g., $\alpha_{ikl}$, $\beta_{ikl}$, and $\tau_{ikl}$ in Section \ref{subsec: mdmdst}). Because components of $\boldsymbol \theta_{ik}$ may lie on different supports (e.g., strictly positive, confined to $[0,1]$, or on the real line), we apply a componentwise link function $\boldsymbol g_k:\mathrm{dom}(\boldsymbol \theta_{ik})\to\mathbb{R}^{d_k}$ (e.g., logarithm, logit, or identity). We then posit a generalized latent–factor model to capture cross–subject and cross–task dependence:
\begin{equation}
\label{eq:lf}
\boldsymbol g_k\!\big(\boldsymbol \theta_{ik}\big)
= \boldsymbol{\mu}_{k} \;+\; \boldsymbol \Gamma_{k}^{\top}\boldsymbol{X}_i \;+\; \boldsymbol \Psi_{k}^{\top}\boldsymbol f_i,
\end{equation}
where $\boldsymbol{\mu}_{k}\in\mathbb{R}^{d_k}$ is a task-specific intercept, $\boldsymbol \Gamma_{k}\in\mathbb{R}^{p\times d_k}$ captures the effect of baseline covariates $\boldsymbol{X}_i\in\mathbb{R}^p$, and $\boldsymbol \Psi_{k}\in\mathbb{R}^{d_f\times d_k}$ is the loading for $d_f$ subject-level latent factors $\boldsymbol f_i\in\mathbb{R}^{d_f}$ shared across tasks. 
By allowing tasks to share subject-level latent factors $\boldsymbol f_i$, this construction enables information borrowing across tasks.

Moreover, only a subset of $\boldsymbol \theta_{ik}$ is intended to be shared across tasks, while the remaining components are inherently task-specific (or unlikely to benefit from sharing). Accordingly, we partition $\boldsymbol \theta_{ik}=\big(\boldsymbol \theta_{ik1}^\top,\,\boldsymbol \theta_{ik2}^{\top}\big)^\top$, where $\boldsymbol \theta_{ik1}$ collects the components shared across tasks and $\boldsymbol \theta_{ik2}$ represents the task-specific components. Although the shared components may differ across tasks, they are expected to be correlated, reflecting common cognitive mechanisms. The choice of shared parameters $\boldsymbol \theta_{ik1}$ versus task-specific parameters $\boldsymbol \theta_{ik2}$ is typically guided by domain knowledge and modeling considerations. In both simulation studies and application in this paper, we treat $(\alpha_{ik0}, \alpha_{ik1}, \beta_{ik1})$ as shared across tasks, whereas the remaining parameters are task-specific, with no latent sharing. 

To encode this structure in \eqref{eq:lf}, we restrict the factor loadings so that the subject–level factors act only on the shared block (i.e., all columns of $\boldsymbol \Psi_{k}$ corresponding to $\boldsymbol \theta_{ik2}$ are set to zero). This prevents spurious sharing through the latent factors while preserving cross–task borrowing for the shared components. Define $\boldsymbol{g}_{i}
= (\boldsymbol{g}_1(\boldsymbol \theta_{i1})^{\top},\,\dots,\,\boldsymbol{g}_K(\boldsymbol \theta_{iK})^{\top})^{\top}$. The cross-task latent-factor representation is
\begin{equation}
\label{eq: lf}
\boldsymbol{g}_{i} \;=\; \boldsymbol{\mu} + \boldsymbol \Gamma^{\top}\boldsymbol{X}_i \;+\; \boldsymbol \Psi^{\top} \boldsymbol f_i,
\end{equation}
where $\boldsymbol{\mu} = (\boldsymbol{\mu}_{1}^{\top}, \dots, \boldsymbol{\mu}_{K}^{\top})^{\top}$, $\boldsymbol \Gamma = (\boldsymbol \Gamma_{1}, \dots, \boldsymbol \Gamma_{K})$, and $\boldsymbol \Psi = (\boldsymbol \Psi_{1}, \dots, \boldsymbol \Psi_{K})$. In Web Appendix B, we show that $\boldsymbol g_i$ in \eqref{eq: lf} is invariant under orthogonal transformations, diagonal rescaling, and translations of the latent factors \(\boldsymbol f_i\), provided that the loading matrix $\boldsymbol \Psi$ and intercept $\boldsymbol \mu$ are adjusted accordingly. Consequently, $(\boldsymbol \mu, \boldsymbol \Psi, \boldsymbol f_i)$ is not identifiable without additional constraints. To address the non-identifiablility issue, we now recover a common identifiable coordinate system
through a post-hoc transformation, including QR decomposition and rescaling of
\(\bm \Psi\), together with centering of \(\boldsymbol f_i\), as described in
Web Appendix F.1. 
This transformation provides an identifiable representation of the model without changing the fitted linear predictors or the likelihood.

\section{Estimation Procedure}\label{sec: ep}
For task $k$, let $\boldsymbol\vartheta_{k1} = \big\{\boldsymbol\mu_{k}, \boldsymbol\Gamma_{k}, \boldsymbol\Psi_{k} \big\}$ denote the DDM-specific parameters, and let $\boldsymbol\vartheta_{k2} = \big\{\pi_{k1}, \gamma_{kl0}, \boldsymbol\gamma_{kl1}\big\}_{l=0}^1$ denote the HMM-specific parameters. Define the full parameter set as $\boldsymbol \Theta = \big\{\boldsymbol\Theta_1, \ldots, \boldsymbol\Theta_K\big\}$, where $\boldsymbol\Theta_k = \big\{\boldsymbol\vartheta_{k1}, \boldsymbol\vartheta_{k2} \big\}$. For subject $i$, let $\boldsymbol U_i = \big\{ U_i^{(1)},\ldots, U_i^{(K)}\big\}$ denote the collection of latent states across tasks. Similarly, let $\boldsymbol{\mathcal D}_i = \big\{\boldsymbol{\mathcal D}_i^{(1)}, \ldots, \boldsymbol{\mathcal D}_i^{(K)}\big\}$ denote the corresponding observed data across tasks. Assume the subject-level latent factors follow a standard normal prior $\boldsymbol f_i \sim p(\boldsymbol f_i) =\mathcal{N}(\boldsymbol 0, \boldsymbol{I}_{d_f})$. Ideally, $\boldsymbol\Theta$ would be estimated by maximum likelihood, marginalizing over the latent variables $(\boldsymbol U_i,\boldsymbol f_i)$ to obtain the observed likelihood for $\boldsymbol{\mathcal D}_i$. In practice, however, this requires evaluating integrals over complex, high-dimensional latent variables of total dimension $\sum^N_{i=1}\sum^K_{k=1}J_{ik}$ with temporal and cross-task dependence, making the observed likelihood analytically intractable and exact computation computationally prohibitive.

To overcome this challenge, consider the joint log-likelihood of the observed data $\boldsymbol{\mathcal D}_i$ and the latent variables $(\boldsymbol U_i,\boldsymbol f_i)$:
\begin{align*}
\mathcal{L}(\boldsymbol\Theta)=&\ \sum_{i=1}^{N}\log p\big(\boldsymbol{\mathcal D}_i,\boldsymbol U_i,\boldsymbol f_i\mid\boldsymbol\Theta\big)\\
=&\ \sum_{i=1}^{N} \sum_{k=1}^{K}\sum^{J_{ik}}_{j=1}\log p\big(\mathcal{D}_{ij}^{(k)}\mid U_{ij}^{(k)},\boldsymbol f_i, \boldsymbol\vartheta_{k1}\big)+\sum_{i=1}^{N}\log p\big(\boldsymbol f_i\big)\\
&+\sum_{i=1}^{N} \sum_{k=1}^{K}\Big\{\log \Pr\big(U_{i1}^{(k)}, \boldsymbol\vartheta_{k2}\big)+\sum^{J_{ik}-1}_{j=1}\log \Pr\big(U_{i,j+1}^{(k)}\mid U_{ij}^{(k)},\boldsymbol\vartheta_{k2}\big)\Big\},
\end{align*}
where $p\big(\mathcal{D}_{ij}^{(k)} \mid U_{ij}^{(k)}, \boldsymbol f_i, \boldsymbol\vartheta_{k1}\big)$ is the DDM emission likelihood in \eqref{eq: fta}, and the last line encodes the HMM initial and transition probabilities. Note that the posterior distribution $p\big(\boldsymbol U_i,\boldsymbol f_i \mid \boldsymbol{\mathcal D}_i\big)$ is intractable; consequently, a direct application of the EM algorithm \citep{dempster1977maximum} is also infeasible. Instead, we approximate the true posterior with a tractable variational surrogate $q\big(\boldsymbol U_i,\boldsymbol f_i\big)$ and optimize it using an EM-VA (EM with VA) algorithm.

\subsection{EM-VA algorithm}
The proposed EM-VA algorithm proceeds iteratively, with iterations indexed by $s=1,2,\ldots$, and terminates when the relative change in evidence lower bound (ELBO), defined later in this section, between successive iterations falls below a prespecified threshold.

The true posterior admits the following conditional factorization: 
$p\big(\boldsymbol U_i,\boldsymbol f_i \mid \boldsymbol{\mathcal D}_i\big) = p\big( \boldsymbol f_i \mid \boldsymbol{\mathcal D}_i\big) p\big( \boldsymbol U_i \mid \boldsymbol{\mathcal D}_i, \boldsymbol f_i \big)$. Guided by this structure, we approximate the first term $p\big( \boldsymbol f_i\mid\boldsymbol{\mathcal D}_i\big)$ with a Gaussian variational family: $q(\boldsymbol f_i)= \mathcal{N}(\boldsymbol m_i, \boldsymbol \Sigma_i)$, where $\boldsymbol m_i=\big(m_{i1}, \ldots,m_{id_f}\big)^\top$ and $\boldsymbol \Sigma_i=\mathrm{diag}\big(\varsigma_{i1}^2,\ldots,\varsigma_{id_f}^2\big)$. For convenience, define $\boldsymbol\varsigma_i=(\varsigma_{i1},\ldots,\varsigma_{id_f})^\top$. Let $\boldsymbol\eta=\{\boldsymbol\eta_i\}_{i=1}^N$ denote all variational parameters with
$\boldsymbol\eta_i=\{\boldsymbol m_i,\boldsymbol\varsigma_i\}$. The second term further factorizes across tasks:
$p\big(\boldsymbol U_i \mid \boldsymbol{\mathcal D}_i,\boldsymbol f_i\big)=\prod^K_{k=1}p\big(\boldsymbol U_i^{(k)}\mid\boldsymbol{\mathcal D}_i^{(k)},\boldsymbol f_i\big)$, and we approximate each factor by $q\big(\boldsymbol U_i^{(k)}\mid \boldsymbol{\mathcal D}_i^{(k)},\boldsymbol{f}_i^{(s-1)}, \boldsymbol \theta_k^{(s-1)} \big)$, where $\boldsymbol{f}_i^{(s-1)}$ is a point estimate of $\boldsymbol f_i$ and $\boldsymbol \theta_k^{(s-1)}$ collects the parameter estimates for task $k$ from iteration $s-1$. This plug-in construction (see Web Appendix C and \cite{guo2026anchored} for a mathematical justification) mirrors the spirit of the EM algorithm: at iteration $s$, the plug-in quantities are computed using the estimates from iteration $s-1$ and are then used to update the current estimates. Importantly, it introduces no additional variational parameters and can be evaluated efficiently over trials via the Baum-Welch algorithm \citep{baum1970maximization}, with details provided in Web Appendix D.

Putting them together, we define the structured VA at the $s$-th iteration as
$q^{(s)}(\boldsymbol U_i,\boldsymbol f_i)=q\big(\boldsymbol f_i\big)\cdot \prod^K_{k=1}q\big(\boldsymbol U_i^{(k)}\mid \boldsymbol{\mathcal D}_i^{(k)},\boldsymbol{f}_i^{(s-1)},\boldsymbol \theta_k^{(s-1)}\big)$, which can be viewed as a partially mean-field approximation. Compared with a fully mean-field approach, which enforces independence between $\boldsymbol f_i$ and $\boldsymbol U_i$, it offers several advantages: it preserves the essential conditional dependence implied by the model (namely, that $\boldsymbol U_i$ depends on $\boldsymbol f_i$ given $\boldsymbol{\mathcal D}_i$), exploits the analytic tractability of the HMM latent states $\boldsymbol U_i$, and introduces fewer variational parameters. Consequently, it reduces variational bias and yields more accurate joint posterior inference.

Under the VA, we choose the surrogate by minimizing the Kullback–Leibler (KL) divergence from $q^{(s)}(\boldsymbol U_i,\boldsymbol f_i)$ to the true posterior: $\sum^N_{i=1}\mathbb{E}_{q^{(s)}(\boldsymbol U_i,\boldsymbol f_i)}\big\{\log q(\boldsymbol U_i,\boldsymbol f_i)\big\}-\sum^N_{i=1}$ $\mathbb{E}_{q^{(s)}(\boldsymbol U_i,\boldsymbol f_i)}\big\{\log p\big(\boldsymbol{\mathcal D}_i,\boldsymbol U_i,\boldsymbol f_i\big)\big\}+\sum^N_{i=1}\log p\big(\boldsymbol{\mathcal D}_i\big)$.
Since the marginal likelihood $p(\boldsymbol{\mathcal D}_i)$ does not depend on the variational distribution $q(\boldsymbol U_i,\boldsymbol f_i)$ and is typically intractable, we instead maximize the ELBO: $\sum^N_{i=1}\mathbb{E}_{q^{(s)}(\boldsymbol U_i,\boldsymbol f_i)}\big\{\log p\big(\boldsymbol{\mathcal D}_i,\boldsymbol U_i,\boldsymbol f_i\big)\big\}-\sum^N_{i=1}\mathbb{E}_{q^{(s)}(\boldsymbol U_i,\boldsymbol f_i)}\big\{\log q(\boldsymbol U_i,\boldsymbol f_i)\big\}$, which is optimized jointly over the model parameters $\boldsymbol\Theta$ and the variational parameters $\boldsymbol\eta$. 
Specifically, at iteration $s$, the ELBO decomposes over subjects as  $\text{ELBO}^{(s)}(\boldsymbol\Theta,\boldsymbol\eta; \boldsymbol{\mathcal D})=\sum_{i=1}^{N} e^{(s)}(\boldsymbol\Theta,\boldsymbol\eta_i; \boldsymbol{\mathcal D}_i)$, with
\begin{align*}
e^{(s)}(\boldsymbol\Theta,\boldsymbol\eta_i; \boldsymbol{\mathcal D}_i)=& \sum_{k=1}^{K}\sum^{J_{ik}}_{j=1}\sum_{l=0}^1\zeta_{ijkl}^{(s)} \mathbb{E}_{q(\boldsymbol f_i; \ \boldsymbol\eta_i)} \left\{ \ell_{ijkl}(\boldsymbol f_i, \boldsymbol\vartheta_{k1}) \right\}+ \sum_{k=1}^{K}\sum_{l=0}^1\zeta_{i1kl}^{(s)}\log\pi_{kl}\\
&+\sum_{k=1}^{K}\sum^{J_{ik}-1}_{j=1}\sum_{l=0}^1\sum_{m=0}^1\xi_{ijklm}^{(s)}\log C_{ijklm}(\boldsymbol\vartheta_{k2}) -\tfrac{1}{2}\sum^{d_f}_{l=1} m_{il}^2+\sum^{d_f}_{l=1}\log\varsigma_{il}-\tfrac{1}{2}\sum^{d_f}_{l=1}\varsigma^2_{il}.
\end{align*}
Here, 
$\zeta_{ijkl}^{(s)}=\Pr\big(U_{ij}^{(k)}=l\mid \boldsymbol{\mathcal D}_i^{(k)},\boldsymbol{f}_i^{(s-1)},\boldsymbol \theta^{(s-1)}\big)$ is the marginal posterior for the latent strategy at a single trial, $\xi_{ijklm}^{(s)}=\Pr\big(U_{i,j+1}^{(k)}=m,U_{ij}^{(k)}=l\mid \boldsymbol{\mathcal D}_i^{(k)}, \boldsymbol{f}_i^{(s-1)},\boldsymbol \theta^{(s-1)} \big)$ is the joint posterior for two consecutive latent strategies, $C_{ijklm}(\boldsymbol\vartheta_{k2}) = \Pr\big(U_{i,j+1}^{(k)}=m\mid U_{ij}^{(k)}=l, \boldsymbol\vartheta_{k2} \big)$ is the transition probability from strategy $l$ to $m$, 
and $\ell_{ijkl}(\boldsymbol f_i, \boldsymbol\vartheta_{k1}) = \log p\big(\mathcal{D}_{ij}^{(k)}\mid U_{ij}^{(k)},\boldsymbol f_i, \boldsymbol\vartheta_{k1}\big)$ is the DDM emission log-likelihood.

Algorithm~\ref{alg1} summarizes the proposed EM–VA procedure. In the E-step (i), we update the point estimate of the latent factor by the posterior mean. 
E-steps (ii)–(vii) outline the forward–backward routine to efficiently compute posteriors over the latent strategy sequences under the HMM structure. The forward variables $a_{ijkl}^{(s)}$ and backward variables $b_{ijkl}^{(s)}$ are defined in Web Appendix D, where full derivations are provided; we omit them here for brevity. We use standard scaling (or log-sum-exp) to avoid numerical underflow. In the M-step, in order to approximate the expectation of the emission log-likelihood, 
we use multivariate Gauss–Hermite quadrature \citep{Jackel2005} and denote it as 
$\widetilde{\ell}_{ijkl}\big(\boldsymbol\vartheta_{k1},\boldsymbol m_i, \boldsymbol\varsigma_i \big)
=\sum_{r=1}^{R} w_r \ 
\ell_{ijkl}\!\left(\boldsymbol m_i+\boldsymbol\Sigma_i^{1/2}\boldsymbol z_r,
\,\boldsymbol\vartheta_{k1}\right) \approx \mathbb{E}_{q(\boldsymbol f_i; \ \boldsymbol\eta_i)} \left( \ell_{ijkl}(\boldsymbol f_i, \boldsymbol\vartheta_{k1}) \right)$, where $\widetilde{\ell}_{ijkl}$ denotes the Gauss–Hermite estimate obtained by a product rule over nodes
$\{\boldsymbol z_r \}_{r=1}^{R}$ with weights $\{w_r\}_{r=1}^{R}$.
Then, we update model parameters via coordinate ascent for the initial state probabilities, HMM-specific parameters, latent factor means, latent factor SDs, and DDM-specific parameters, as described in M-steps (i) – (v), respectively. We alternate updates of the variational parameters $\boldsymbol{\eta}_i$ and the model parameters $\boldsymbol \theta$ until convergence, yielding $\boldsymbol{\eta}_i=\boldsymbol{h}(\boldsymbol \theta, \boldsymbol{\mathcal{D}}_i)$.
This induces a profiled objective function and places the estimation of $\boldsymbol \theta$ within an M-estimation framework \citep{westling2019beyond,zeng2023zero}.
In Web Appendix E, we prove that, under appropriate regularity conditions, the global ELBO maximizer $\widehat{\boldsymbol \theta}$ is consistent; that is, $\widehat{\boldsymbol \theta} \xrightarrow{p} \boldsymbol \theta_*$, where $\boldsymbol \theta_{*}$ is the true population parameter.

\begin{algorithm}[h!]
\small
\caption{EM-VA algorithm}\label{alg1}
\begin{algorithmic}
    \State \textbf{E-step}: for $i=1,\ldots,N$, $k=1,\ldots,K$, and $l=0,1$,
    \State \quad (i) \textbf{Set latent factor}: $\boldsymbol f_i^{(s-1)}\gets$  $\boldsymbol m_i^{(s-1)}$;
    \State \quad (ii) \textbf{Compute HMM transition}: for $j=2,\ldots,J_{ik}$, $C_{i,j-1,k,m,l}^{(s-1)} = C_{i,j-1,k,m,l}(\boldsymbol\vartheta_{k2}^{(s-1)})$;
    \State \quad (iii) \textbf{Compute DDM likelihood}: for $j=1,\ldots,J_{ik}$, $\ell_{ijkl}^{(s-1)}=\ell_{ijkl}(\boldsymbol f_i^{(s-1)}, \boldsymbol\vartheta_{k1}^{(s-1)})$;
    \State \quad (iv) \textbf{Initialize forward and backward variables}: $a_{i1kl}^{(s)}\gets\pi_{kl}^{(s-1)}\ell_{i1kl}^{(s-1)}$ \quad and \quad $b_{iJ_{ik}kl}^{(s)}\gets1$;
    \State \quad (v) \textbf{Forward recursion}: for $j=2,\ldots,J_{ik}$, update  $a_{ijkl}^{(s)}\gets \ell_{ijkl}^{(s-1)}\sum^1_{m=0}a_{i,j-1,k,m}^{(s)}C_{i,j-1,k,m,l}^{(s-1)}$;  
    \State \quad (vi) \textbf{Backward recursion}: for $j=J_{ik}-1,\ldots,1$, update 
    $b_{ijkl}^{(s)}\gets \sum^1_{m=0}b_{i,j+1,k,m}^{(s)}\ell_{i,j+1,k,m}^{(s-1)}C_{ijklm}^{(s-1)}$; 
    \State \quad (vii) \textbf{Posterior weights update}: for $j=1,\ldots,J_{ik}-1$, \begin{align*}\zeta_{ijkl}^{(s)}\gets\frac{a_{ijkl}^{(s)}b_{ijkl}^{(s)}}{\sum^1_{m=0}a_{iJkm}^{(s)}} \quad \text{and} \quad \xi_{ijklm}^{(s)}\gets\frac{a_{ijkl}^{(s)}b_{i,j+1,k,m}^{(s)}\ell_{i,j+1,k,m}^{(s-1)}C_{ijklm}^{(s-1)}}{\sum^1_{m=0}a_{iJkm}^{(s)}}.
    \end{align*}
    \State \textbf{M-step}: for $i=1,\ldots,N$, $k=1,\ldots,K$, and $l=0,1$,
    \State \quad (i) \textbf{Initial state probabilities update}: $\pi_{k1}^{(s)}\gets N^{-1}\sum^N_{i=1}\zeta_{i1k1}^{(s)}$;
    \State \quad (ii) \textbf{HMM-specific parameters update}: 
    \begin{align*}
    \big\{\gamma_{kl0}^{(s)},\boldsymbol\gamma_{kl1}^{(s)}\big\}\gets\argmin_{\gamma_{kl0},\boldsymbol\gamma_{kl1}}\sum^N_{i=1}\sum^{J_{ik}-1}_{j=1}&\Big[-\xi_{ijkl1}^{(s)}\big(\gamma_{kl0}+\boldsymbol\gamma_{kl1}^{\top} \boldsymbol{X}_{i}\big)\\
    &+\big(\xi_{ijkl0}^{(s)}+\xi_{ijkl1}^{(s)}\big)\log\big\{1+\exp\big(\gamma_{kl0}+\boldsymbol\gamma_{kl1}^{\top} \boldsymbol{X}_{i}\big)\big\}\Big];    
    \end{align*}
    \State  \quad (iii) \textbf{Latent factor mean update}: 
    \begin{align*}
    \boldsymbol m_i^{(s)}\gets\argmin_{\boldsymbol m_i} \Big[\sum_{k=1}^{K}\sum^{J_{ik}}_{j=1}\sum_{l=0}^1-\zeta_{ijkl}^{(s)}\widetilde{\ell}_{ijkl}\big(\boldsymbol\vartheta_{k1}^{(s-1)}, \boldsymbol m_i, \boldsymbol\varsigma_i^{(s-1)}\big)+ \tfrac{1}{2}\sum^{d_f}_{l=1}m_{il}^2\Big];
    \end{align*}
    \State  \quad (iv) \textbf{Latent factor SD update}:
    \begin{align*}\boldsymbol\varsigma^{(s)}_i\gets\argmin_{\boldsymbol\varsigma_i}\Big[\sum_{k=1}^{K}\sum^{J_{ik}}_{j=1}\sum_{l=0}^1-\zeta_{ijkl}^{(s)}\widetilde{\ell}_{ijkl}\big(\boldsymbol\vartheta_{k1}^{(s-1)},\boldsymbol m_i^{(s)}, \boldsymbol\varsigma_i\big)+\tfrac{1}{2}\sum^{d_f}_{l=1}\varsigma_{il}^2-\sum^{d_f}_{l=1}\log\varsigma_{il}\Big];
    \end{align*}
    \State \quad (v) \textbf{DDM-specific parameters update}:
    \begin{align*}
    \boldsymbol\vartheta_{k1}^{(s)}\gets\argmin_{\boldsymbol\vartheta_{k1}}\sum_{i=1}^{N}\sum^{J_{ik}}_{j=1}\sum_{l=0}^1-\zeta_{ijkl}^{(s)}\widetilde{\ell}_{ijkl}\big(\boldsymbol\vartheta_{k1},\boldsymbol m_i^{(s)}, \boldsymbol\varsigma_i^{(s)}\big).
    \end{align*}
\end{algorithmic}
\end{algorithm}

Due to the potential non-concavity of the optimization problem and the use of multivariate Gauss–Hermite quadrature, the ELBO may not be strictly monotone, and multiple local maxima may exist; therefore, the algorithm may not be guaranteed to attain the global optimum. In our numerical studies, we observe that the ELBO sequence is generally increasing and stabilizes as the algorithm converges, with only negligible fluctuations, and the algorithm typically converges within a reasonable number of iterations. 

\section{Simulation Studies}\label{sec: ss}
This section evaluates the finite-sample performance of the proposed method. For each setting described below, we perform $100$ Monte Carlo replications with sample sizes $N\in\{100,200\}$. In Web appendix F.2, we show that 100 replicates are sufficient to provide accurate point estimates.

\subsection{Simulation design and data generation}
We consider two settings. The first setting uses parameter values similar to those estimated from the real data that motivated the study, and the second setting is identical except that there is no sharing of parameters across tasks (i.e., the factor loadings are set to zero). In both settings, we consider $K=2$ behavioral tasks: a PRT ($k=1$) with $J_1=100$ trials and a FT ($k=2$) with $J_2=70$ trials. In the PRT, the stimuli sequence $\{S_{ij}^{(1)}\}_{j=1}^{J_1}$ is i.i.d. $\operatorname{Bernoulli}(0.5)$, while in the FT, $\{ S_{ij}^{(2)} \}_{j=1}^{J_2}$ is i.i.d. $\operatorname{Bernoulli}(0.65)$. 
For $k\in\{1,2\}$, the latent strategy sequence \(\{U_{ij}^{(k)}\}_{j=1}^{J_k}\) follows a two-state HMM as specified in \eqref{eq: hmm}. 
The HMM transition probabilities depend on a single baseline covariate \(X_i \sim \operatorname{Bernoulli}(0.7)\), sampled independently across subjects.
Conditional on $S_{ij}^{(k)}$ and $U_{ij}^{(k)}$, the action $A_{ij}^{(k)}$ and RT $T_{ij}^{(k)}$ are generated from a mixture of DDMs, as specified in \eqref{eq: fta}.
Reward outcomes $R_{ij}^{(1)}$ in the PRT are determined by the stimuli–action pair at trial $j$. Incorrect choices yield $R_{ij}^{(1)}=0$. For correct choices, feedback is probabilistic: on rich trials, $R_{ij}^{(1)}\sim \operatorname{Bernoulli}(0.75)$; on lean trials, $R_{ij}^{(1)}\sim \operatorname{Bernoulli}(0.30)$.

The subject–task parameters $\boldsymbol \theta_{ik}$ are assumed to follow the latent factor model $\boldsymbol g_k(\boldsymbol \theta_{ik}) = \boldsymbol{\mu}_{k} + \boldsymbol \Psi_{k}^{\top}\boldsymbol f_i$ with $d_f=2$ latent factors. The columns of $\boldsymbol \Psi_{k}$ corresponding to task-specific parameters are structurally zero and are therefore omitted from the notation. Consequently, $\boldsymbol{\mu}_k$ represents the mean vector for the shared-parameters only, while task-specific parameters are introduced using separate notation. We specify the link functions as $\boldsymbol g_1=\boldsymbol g_2=(\log,\ \log,\ \operatorname{logit})$, applied componentwise to the three shared parameters. The initial reward expectation in the PRT is fixed as $Q_{i1}=\big(\begin{smallmatrix}
0.5 & 0\\
0 & 0.5
\end{smallmatrix}\big)$ and is common to both settings. 

The common parameters across both settings are as follows. For the HMM: in the PRT, the initial engagement probability is $\pi_{11}=0.95$, with transition parameters $\gamma_{100}=\gamma_{101}=-0.3$ and $\gamma_{110}=\gamma_{111}=1.3$; in the FT, the initial focused probability is $\pi_{21}=0.8$, with transition parameters $\gamma_{200}=\gamma_{201}=-0.6$ and $\gamma_{210}=\gamma_{211}=1.2$. For the shared DDM parameters, the population means are $\boldsymbol\mu_{1}=(\log0.4, \log1.5, \log1.2)^\top$ and $\boldsymbol\mu_{2}=(\log1.1, \log2, \log0.5)^\top$. Task-specific DDM parameters are: in the PRT, learning rate $b=0.03$, scaling parameter $c=2.5$, and non-decision time \(\tau_{ik1}=0.14\); in the FT, controlled drift $v_c=3$, automatic drift $v_a=1.5$, attenuation $\rho=0.1$, and non-decision time \(\tau_{ik2}=0.14\).

The two settings differ only in the factor loading matrices. In Setting 1 (with sharing), the factor loadings are $\boldsymbol\Psi_{1}=\big(\begin{smallmatrix}
0.1 & 0.2 & -0.1\\
0 & 0.1 & -0.1
\end{smallmatrix}\big)$
and $\boldsymbol\Psi_{2}=\big(\begin{smallmatrix}
0.1 & 0.15 & -0.1\\
-0.15 & -0.1 & 0.1
\end{smallmatrix}\big)$, while in Setting 2 (no sharing), the factor loadings are set to zero:
$\boldsymbol\Psi_{1}=\boldsymbol\Psi_{2}=\big(\begin{smallmatrix}
0 & 0 & 0\\
0 & 0 & 0
\end{smallmatrix}\big)$.

\subsection{Simulation results}
We compare our proposed joint modeling method, SHIFT, with the separate task modeling method, SPLIT, introduced in Section~\ref{subsec: mdmdst}.
The relative bias (RB) and empirical standard error (ESE) for the task-specific parameters, based on $100$ replicates, are presented in Table \ref{tab1}. Overall, Table \ref{tab1} shows that under Setting 1 (with sharing), SHIFT consistently yields more accurate and efficient estimates across parameters than SPLIT. In contrast, under Setting 2 (no sharing), SHIFT performs comparably to SPLIT (the true model in this setting), with SPLIT being slightly more efficient. Specifically, in Setting 1, SPLIT severely underestimates the learning rate $b$ in the PRT, with an absolute RB over $19$\%. For the FT, SPLIT underestimates both the controlled drift $v_c$ and the automatic drift $v_a$, with absolute RBs exceeding $12$\% and $13$\%, respectively, and provides a highly inaccurate estimate of the attenuation factor $\rho$, with RB over $250$\%. Additionally, SPLIT performs particularly poorly in estimating the HMM transition parameters, exhibiting large absolute RB values. 

\begin{table}[h!]
    \centering
    \caption{Summary of the PRT and FT parameter estimates over 100 simulations.}
    \label{tab1}
    \resizebox{\textwidth}{!}{\begin{tabular}{c c r c r r c r r c r r c}
    \toprule
    & & \multicolumn{5}{c}{Setting 1 (with sharing)} & & \multicolumn{5}{c}{Setting 2 (no sharing)}\\
    \cmidrule{3-7}\cmidrule{9-13}
    & & \multicolumn{2}{c}{SHIFT} & & \multicolumn{2}{c}{SPLIT} & & \multicolumn{2}{c}{SHIFT} & & \multicolumn{2}{c}{SPLIT}\\
    \cmidrule{3-4}\cmidrule{6-7}\cmidrule{9-10}\cmidrule{12-13}
    $N$ & Parameters & \multicolumn{1}{c}{RB} & ESE & & \multicolumn{1}{c}{RB} & ESE & & \multicolumn{1}{c}{RB} & ESE & & \multicolumn{1}{c}{RB} & ESE\\
    \midrule
    \multicolumn{13}{c}{PRT}\\
    \midrule
    100 & $b$ & -0.011 & 0.003 & & -0.198 & 0.003 & & 0.009 & 0.004 & & -0.009 & 0.003\\
    & $c$ & -0.002 & 0.038 & & -0.016 & 0.040 & & 0.009 & 0.043 & & 0.028 & 0.034\\
    & $\tau_1$ & -0.001 & 0.000 & & -0.013 & 0.000 & & -0.001 & 0.000 & & -0.001 & 0.000\\
    & $\pi_{11}$ & 0.005 & 0.022 & & -0.024 & 0.029 & & 0.007 & 0.020 & & 0.006 & 0.020\\
    & $\gamma_{100}$ & -0.003 & 0.071 & & 0.440 & 0.086 & & 0.010 & 0.067 & & 0.026 & 0.067\\
    & $\gamma_{101}$ & 0.024 & 0.093 & & -0.356 & 0.107 & & -0.016 & 0.095 & & -0.036 & 0.095\\
    & $\gamma_{110}$ & 0.001 & 0.058 & & -0.106 & 0.083 & & 0.001 & 0.064 & & -0.005 & 0.064\\
    & $\gamma_{111}$ & 0.005 & 0.088 & & -0.090 & 0.126 & & 0.009 & 0.094 & & 0.008 & 0.094\\
    200 & $b$ & -0.029 & 0.003 & & -0.202 & 0.003 & & 0.019 & 0.002 & & 0.001 & 0.002\\
    & $c$ & -0.000 & 0.036 & & -0.013 & 0.032 & & 0.009 & 0.039 & & 0.026 & 0.034\\
    & $\tau_1$ & -0.001 & 0.000 & & -0.011 & 0.000 & & -0.000 & 0.000 & & -0.002 & 0.000\\
    & $\pi_{11}$ & 0.001 & 0.019 & & -0.027 & 0.023 & & 0.002 & 0.018 & & 0.001 & 0.019\\
    & $\gamma_{100}$ & 0.007 & 0.050 & & 0.401 & 0.055 & & -0.008 & 0.050 & & 0.012 & 0.050\\
    & $\gamma_{101}$ & -0.003 & 0.068 & & -0.370 & 0.067 & & 0.010 & 0.073 & & -0.012 & 0.072\\
    & $\gamma_{110}$ & -0.003 & 0.045 & & -0.100 & 0.055 & & -0.002 & 0.044 & & -0.008 & 0.044\\
    & $\gamma_{111}$ & 0.000 & 0.061 & & -0.089 & 0.079 & & 0.005 & 0.057 & & 0.005 & 0.056\\
    \midrule
    \multicolumn{13}{c}{FT}\\
    \midrule
    100 & $v_c$ & -0.006 & 0.063 & & -0.120 & 0.078 & & -0.003 & 0.049 & & -0.002 & 0.047\\
    & $v_a$ & -0.002 & 0.033 & & -0.136 & 0.041 & & -0.000 & 0.025 & & -0.000 & 0.025\\
    & $\rho$ & -0.022 & 0.026 & & 2.533 & 0.134 & & -0.024 & 0.027 & & -0.019 & 0.027\\
    & $\tau_2$ & -0.000 & 0.001 & & -0.018 & 0.001 & & 0.001 & 0.001 & & 0.001 & 0.001\\
    & $\pi_{21}$ & -0.008 & 0.061 & & -0.123 & 0.088 & & 0.002 & 0.056 & & 0.001 & 0.057\\
    & $\gamma_{200}$ & -0.068 & 0.154 & & 1.046 & 0.440 & & -0.052 & 0.155 & & -0.049 & 0.155\\
    & $\gamma_{201}$ & 0.087 & 0.169 & & -0.174 & 0.313 & & 0.070 & 0.180 & & 0.065 & 0.180\\
    & $\gamma_{210}$ & -0.003 & 0.120 & & 0.103 & 0.242 & & 0.005 & 0.108 & & 0.002 & 0.105\\
    & $\gamma_{211}$ & 0.018 & 0.151 & & -0.119 & 0.222 & & -0.000 & 0.140 & & -0.001 & 0.140\\
    200 & $v_c$ & -0.014 & 0.060 & & -0.121 & 0.049 & & 0.001 & 0.037 & & 0.002 & 0.034\\
    & $v_a$ & -0.010 & 0.030 & &-0.141 & 0.026 & & 0.000 & 0.017 & & 0.001 & 0.017\\
    & $\rho$ & -0.010 & 0.019 & & 2.514 & 0.089 & & -0.002 & 0.022 & & 0.005 & 0.021\\
    & $\tau_2$ & -0.002 & 0.001 & & -0.018 & 0.001 & & 0.000 & 0.001 & & -0.000 & 0.001\\
    & $\pi_{21}$ & 0.011 & 0.045 & & -0.119 & 0.052 & & -0.008 & 0.038 & & -0.008 & 0.038\\
    & $\gamma_{200}$ & -0.014 & 0.112 & & 1.025 & 0.278 & & -0.010 & 0.091 & & -0.006 & 0.091\\
    & $\gamma_{201}$ & 0.034 & 0.124 & & -0.213 & 0.218 & & 0.002 & 0.104 & & -0.003 & 0.104\\
    & $\gamma_{210}$ & 0.029 & 0.105 & & 0.084 & 0.179 & & -0.008 & 0.093 & & -0.011 & 0.090\\
    & $\gamma_{211}$ & 0.004 & 0.126 & & -0.114 & 0.185 & & 0.010 & 0.099 & & 0.010 & 0.100\\
    \bottomrule
    \multicolumn{12}{l}{\footnotesize (RB): relative bias; (ESE): empirical standard error.}
    \end{tabular}}
\end{table}

\begin{figure}[h!]
    \begin{subfigure}{\textwidth}
    \centering
    \includegraphics[width=\linewidth]{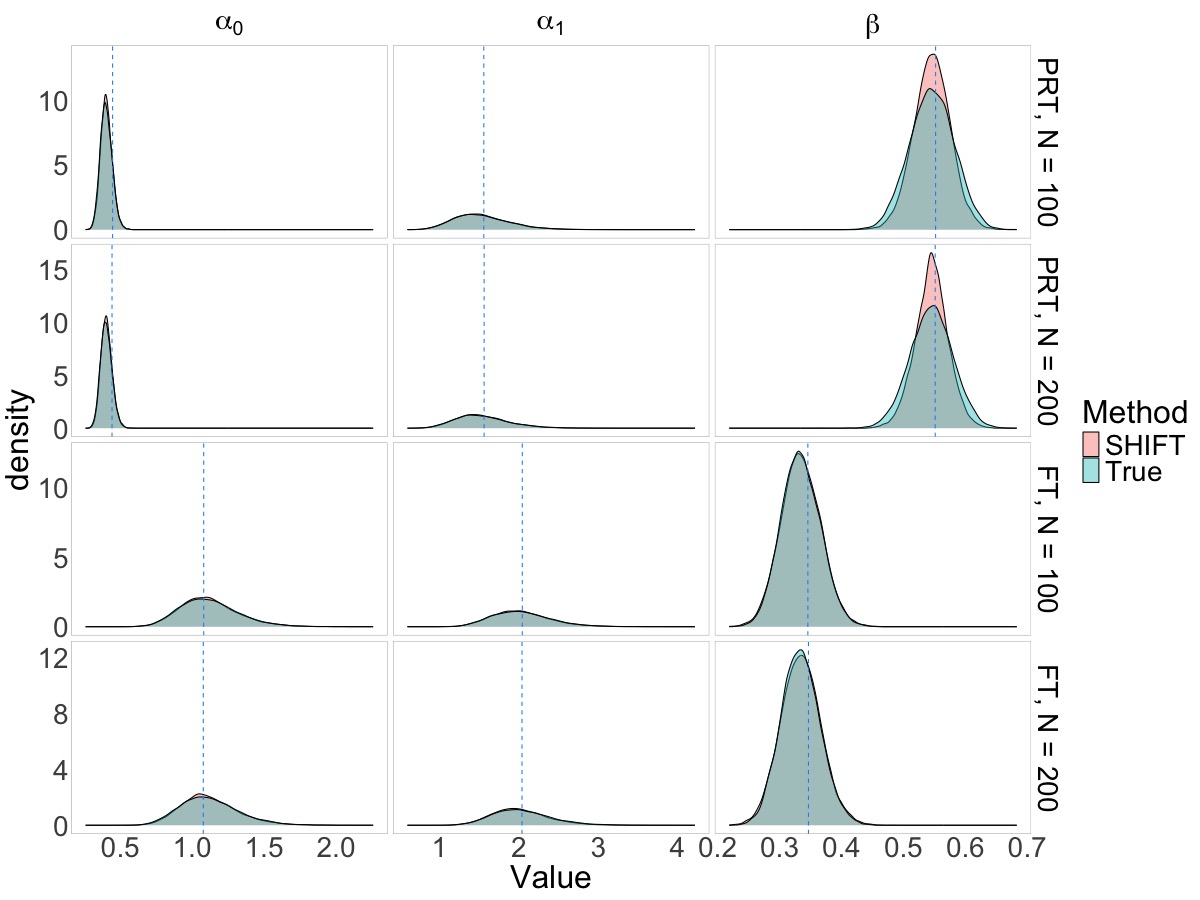}
    \caption{Density plots of shared parameters and their SHIFT estimates, where the blue lines represent the means of the SPLIT estimates.}
    \label{Fig2a}
    \end{subfigure}
    \begin{subfigure}{\textwidth}
    \centering
    \includegraphics[width=\linewidth]{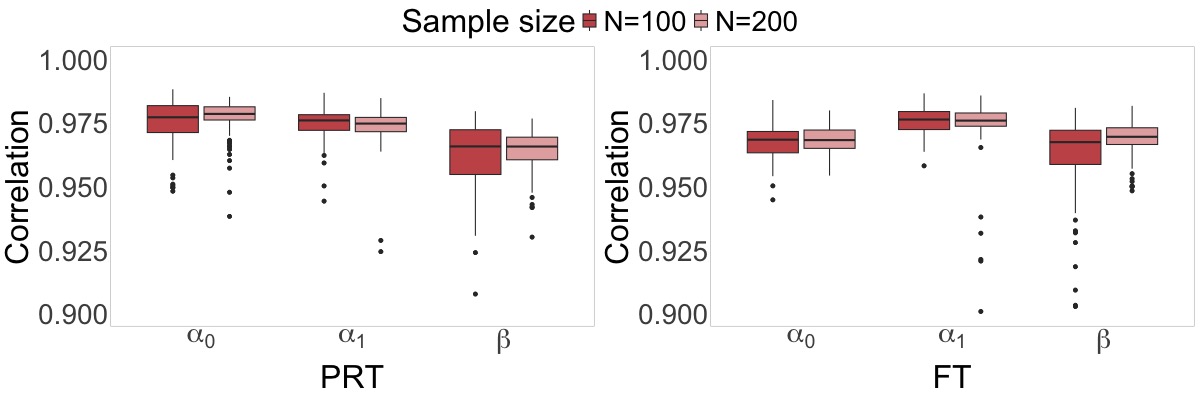}
    \caption{Correlations between the true shared parameters and their SHIFT estimates.}
    \label{Fig2b}
    \end{subfigure}
    \caption{Density distributions of the true and SHIFT estimated shared parameters and their correlations, Setting 1.}
\end{figure}

Notably, SHIFT is capable of estimating the subject-specific shared parameters, whereas SPLIT is limited to estimating their population-level counterparts. Figure~\ref{Fig2a} displays the kernel density estimates of the true and estimated shared parameters for Setting 1. SHIFT captures the distribution of the shared parameters, whereas the SPLIT estimates exhibit mean bias relative to the truth. In Setting 2, such shared structure does not exist by design; therefore, no corresponding plot is shown. Nevertheless, SHIFT is able to accurately estimate $\alpha_0$, $\alpha_1$, and $\beta$, with RB less than $0.003$ under Setting 2.  
For each replicate, we also compute the correlation across subjects between the true shared parameters and their SHIFT estimates under Setting 1. Figure~\ref{Fig2b} presents these correlations, most of which exceed $90\%$, highlighting SHIFT's accuracy in recovering subject-specific shared effects. A small number of extreme outliers are omitted from the figure to enhance clarity.

\begin{figure}[h!]
    \begin{subfigure}{\textwidth}
    \centering
    \includegraphics[width=\linewidth]{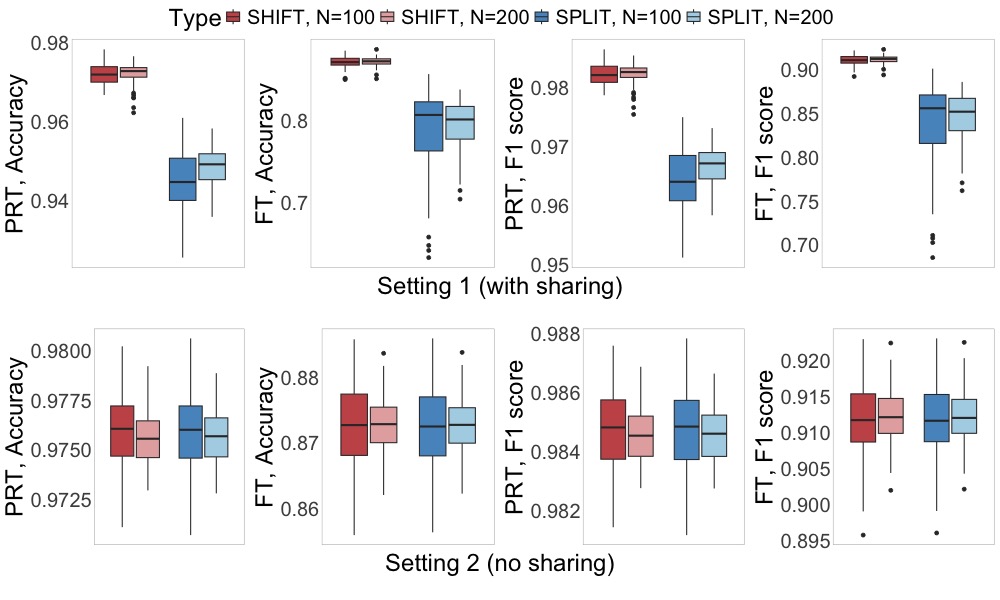}
    \caption{Estimation accuracy and F1 score for the latent state $U_{ij}^{(k)}$.}
    \vspace{8mm}
    \label{Fig3a}
    \end{subfigure}
    \begin{subfigure}{\textwidth}
    \centering
    \includegraphics[width=\linewidth]{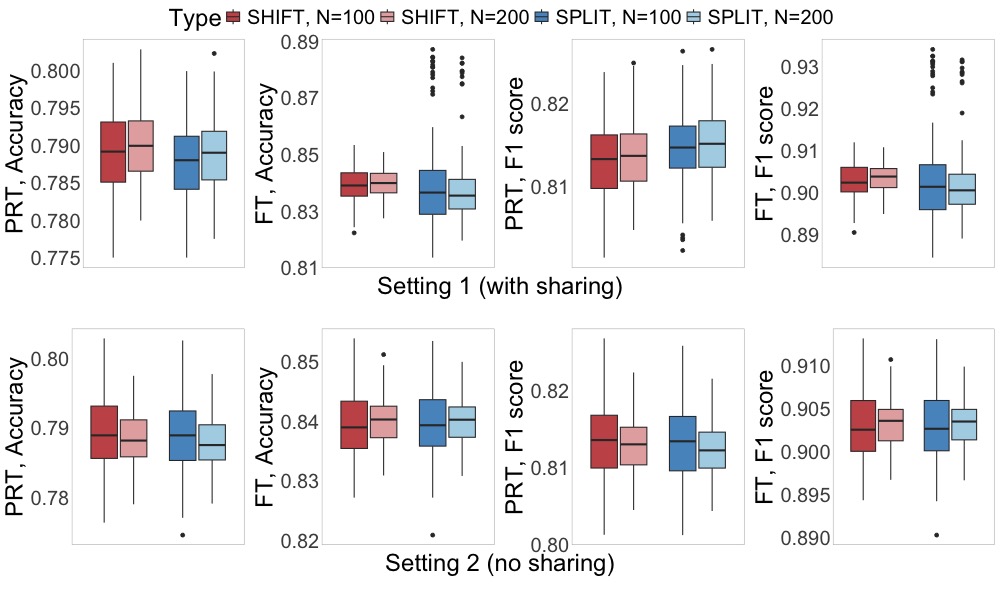}
    \caption{Estimation accuracy and F1 score for the actions $A_{ij}^{(k)}$.}
    \label{Fig3b}
    \end{subfigure}
    \caption{Estimation accuracy and F1 score for the latent state $U_{ij}^{(k)}$ and actions $A_{ij}^{(k)}$.}
\end{figure}

We estimate the latent decision-making strategies $U_{ij}^{(k)}$ by computing their posterior probabilities and applying a $0.5$ hard threshold to obtain estimates, and we assess performance using accuracy and F1 scores, as shown in Figure \ref{Fig3a}. Across both tasks, SHIFT outperforms SPLIT on both metrics under Setting 1, indicating more reliable recovery of latent strategies when sharing exists; and SHIFT yields similar results to SPLIT on both metrics under Setting 2, where SPLIT is the true model. Notably, SHIFT and SPLIT share the same HMM component; the gains arise from SHIFT’s more accurate estimation of DDM parameters, which in turn improves HMM inference.
Additionally, Figure \ref{Fig3b} displays the estimation accuracy for actions $A_{ij}^{(k)}$; in most cases, SHIFT attains higher accuracy and F1 than SPLIT, indicating a more accurate reconstruction of the decision-making process. Moreover, performance generally improves with larger sample sizes.

For simulations with sample size $100$, each E-step takes less than one second, while each M-step requires approximately 13 minutes on a computing cluster using 5 cores and 2 GB of memory per node. Increasing the number of cores to $10$ yields similar run time for sample size $200$. Additional simulation results, including post-hoc transformations for identifiability, a setting without strategy switching, a setting with error-contaminated RTs, and a setting with non-Markovian strategy switching are summarized in Web Appendix F. These results demonstrate that SHIFT is quite robust to model misspecification and measurement error.

\section{Application to EMBARC Study}\label{sec: aes}
Preliminary analyses suggest that behavioral patterns may differ across blocks. Therefore, we focus on the first block (100 trials) of each PRT session and the first block (70 trials) of each FT session. A total of 183 subjects completed both tasks: 31 in the CTL group and 152 in the MDD group. We define $X_i=0$ for CTL subjects and $X_i=1$ for MDD subjects. Following standard practice \citep{huys2013mapping}, RTs shorter than 150 milliseconds (ms) were truncated to 150 ms, and RTs longer than 1500 ms were truncated to 1500 ms.

We analyze the two tasks using SHIFT with multiple random initializations, selecting the run that yields the largest ELBO. The ELBO trajectories for 10 different initializations are shown in Web Figure G.4. These trajectories demonstrate that different starting values can converge to different ELBO values, indicating the presence of multiple local optima. As noted in \cite{blei2017variational}, this is not necessarily a disadvantage: mixture models typically have many posterior modes due to label switching, and recovering one such mode is often sufficient for exploring latent clusters or predicting new observations.

For the first trial, the estimated probability of being engaged is $99.9$\% in the PRT, and the estimated probability of being focused is $88.9$\% in the FT. The higher probability of engagement and focus in the PRT may be attributable to reward incentives. Within the PRT, CTL subjects exhibit a $99.7$\% estimated probability of remaining in the engaged state if they were engaged in the previous trial, while MDD subjects have a slightly lower estimated probability of $99.0$\%. Conversely, if subjects were in the lapsed state during the previous trial, the estimated probability of transitioning to the engaged state is $99.9$\% for CTL and a much smaller value, $59.0$\%, for MDD. Within the FT, the estimated transition probabilities are as follows: $99.5$\% for CTL staying focused, $99.1$\% for MDD staying focused, $86.3$\% for CTL transitioning from a reduced state, and $31.1$\% for MDD transitioning from a reduced state. 

In the PRT, the estimated correct decision rates are $77.2$\% in the estimated engaged state and $51.2$\% in the estimated lapsed state for MDD, compared with $80.4$\% and $36.4$\%, respectively, for CTL. In the FT, the corresponding rates are $92.5$\% in the estimated focused state and $19.5$\% in the estimated reduced state for MDD, versus $92.6$\% and $14.3$\% for CTL. These findings show that MDD exhibits a modestly lower correct decision rates than CTL in the estimated engaged, and focused states. In the estimated lapsed state, performance in both groups approach the level of random guessing.

\begin{figure}[h!]
    \centering
    \includegraphics[width=\linewidth]{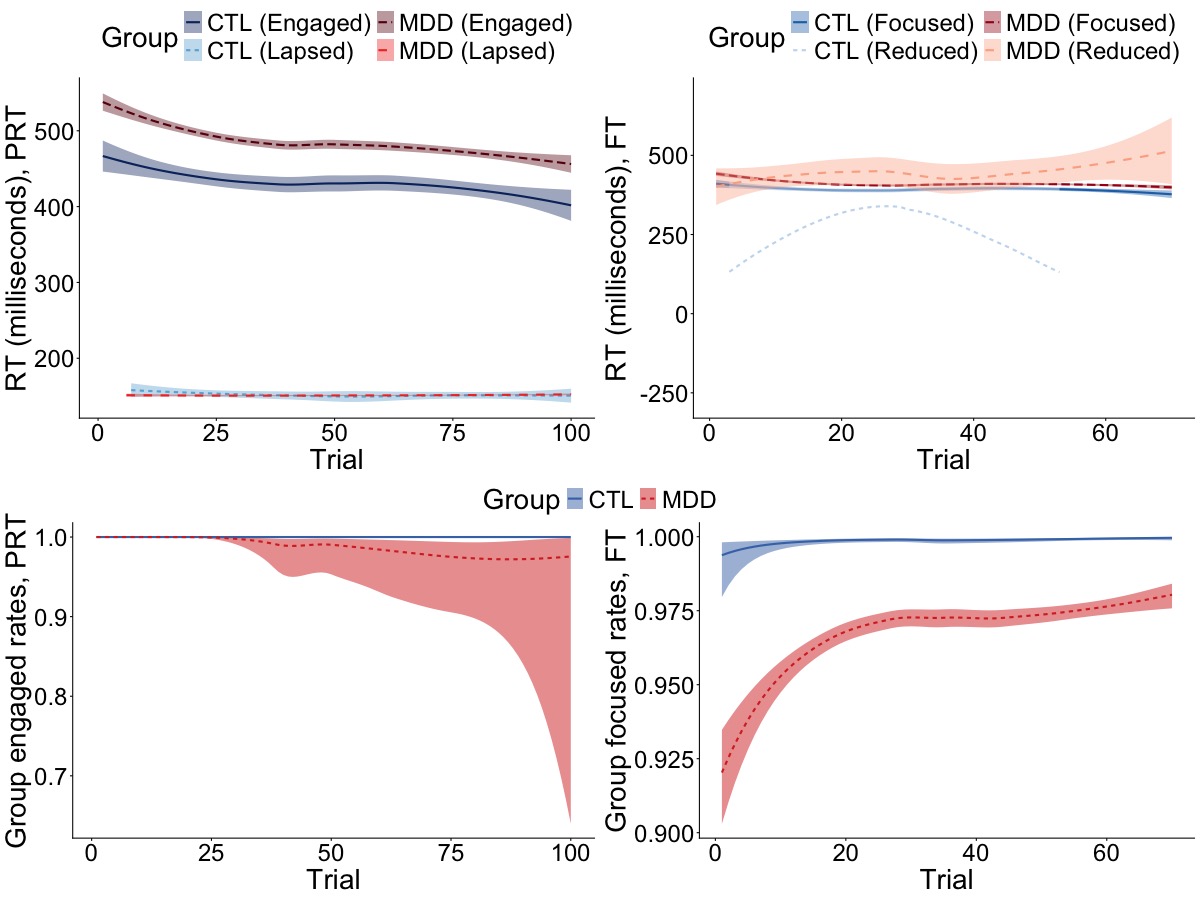}
    \caption{Smoothed RT trajectories for CTL and MDD in the estimated engaged/focused and lapsed/reduced states across the PRT and FT (top), and estimated group-level engaged/focused tendencies for CTL and MDD across both tasks (bottom). Smoothed curves were obtained by fitting locally weighted regression (LOESS) models separately for each group; shaded ribbons indicate 95\% pointwise confidence bands around the mean curves, computed under a local regression framework using the standard errors of the fitted curves and a normal approximation. We omit the confidence interval for CTL (Reduced) due to limit observations in this group.}
    \label{Fig4}
\end{figure}

Figure~\ref{Fig4} shows the smoothed RT trajectories for each estimated decision-making strategy, separated by group (CTL and MDD), in the top panel. The bottom panel displays the average group-level estimated engagement rates (defined as the mean posterior probability of being in the engaged mode) for the PRT and the analogous focused rates for the FT, at each trial. Across both tasks, individuals with MDD exhibited longer average decision times than those in the CTL group in the engaged, focused, and reduced modes. In the lapsed mode, which corresponds to random guessing, both groups showed similarly short RTs, substantially faster than in the engaged mode. Additionally, CTL participants were more likely to be in the engaged state during the PRT and in the focused state during the FT, consistent with more deliberate decision-making. In contrast, individuals with MDD showed a greater tendency toward lapsed (careless) decision-making across both tasks. Engagement and focus rates remained stable across trials in the CTL group for both tasks. In contrast, the MDD group’s engagement rate in the PRT steadily declined across trials, while their focused rate in the FT increased slightly during the first half of the trials before leveling off.

\begin{figure}[h!]
    \begin{subfigure}{\textwidth}
         \centering
         \includegraphics[width=\textwidth]{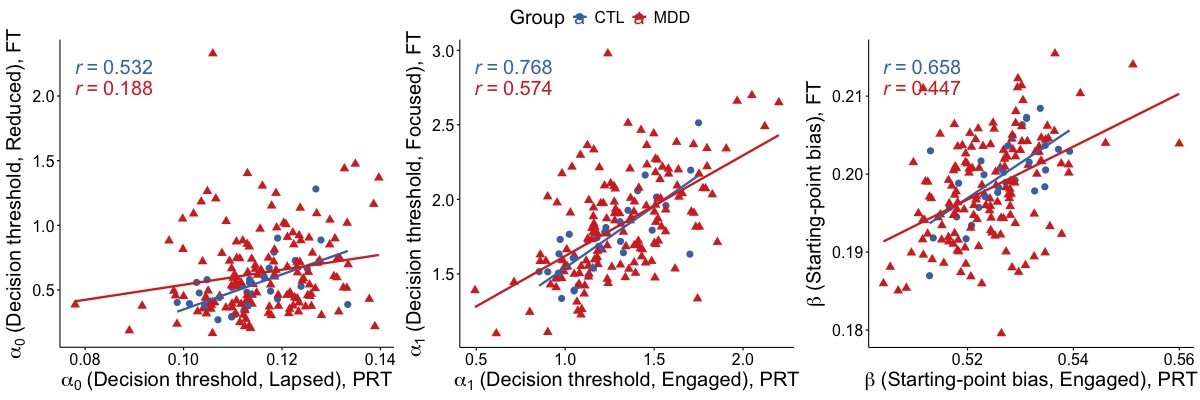}
         \caption{Scatter plots of the estimated shared parameters across tasks, with each point representing an individual subject.}
         \label{Fig5a}
    \end{subfigure}
    \begin{subfigure}{\textwidth}
         \centering
         \includegraphics[width=\textwidth]{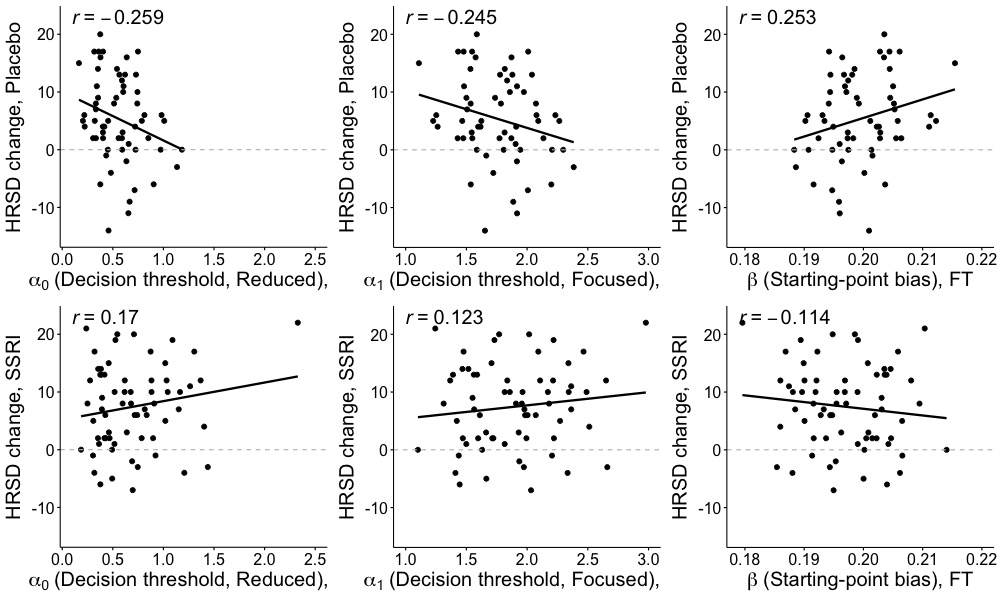}
         \caption{Scatter plots of HRSD change versus estimated shared parameters for the FT, with each point representing an individual subject. The top panel shows subjects receiving placebo and the bottom panel shows subjects receiving SSRI.}
         \label{Fig5b}
    \end{subfigure}
    \caption{Scatter plots of the estimated shared parameters.}
\end{figure}

The estimated shared parameters between the PRT and FT are shown in Figure~\ref{Fig5a}. All three parameters displayed moderate correlations, which likely account for the behavioral correlation between the PRT and FT reported in \cite{bian2025ddm}, as well as the RT correlations observed in Figure~\ref{Fig1c}. In subsequent analyses, the shared parameters for the FT showed suggestive treatment‐modulation patterns ($p$-values less than 0.05, with confidence intervals excluding zero), highlighting their potential as novel behavioral markers of therapeutic outcomes. Figure~\ref{Fig5b} presents scatter plots of treatment response against these parameters. Although the statistical evidence is modest, the direction and magnitude of the effects suggest that these computational parameters may hold promise as objective markers of antidepressant response. We emphasize that these results are exploratory and should be interpreted as hypothesis-generating rather than confirmatory.

We also examined associations between the two latent factors and a broad range of behavioral and neuroimaging measures in the EMBARC study, including behavioral phenotyping, electroencephalography (EEG), functional magnetic resonance imaging (fMRI), diffusion tensor imaging (DTI), and structural MRI. After applying the post hoc transformation described in Web Appendix B and correcting for multiple comparisons, the second latent factor was significantly associated with: (1) the mean RT difference (incongruent minus congruent) in the emotional conflict task fMRI, (2) the BR3–BR9 mean signal in the first resting-state fMRI block, (3) the median correct total RT z-score on the A-not-B task, (4) the median correct RT z-score on the choice reaction time task, and (5) the total number of valid words produced in the word fluency task. To investigate interpretability, we also inspected the transformed $\boldsymbol \Psi$ matrix. The engaged boundary in the PRT primarily loaded on the first latent factor, whereas the reduced and focused boundaries in the FT primarily loaded on the second latent factor.

\section{Discussion}\label{sec: d}
In this paper, we focused on jointly modeling multiple behavioral tasks to improve understanding of cognitive and decision-making processes. The proposed method, SHIFT, relies on the EM-VA algorithm, and may be sensitive to the initializations. For practitioners, we recommend using multiple random starts and then selecting the run with the largest ELBO. When SPLIT is available, its estimates provide an effective warm start for SHIFT.

We applied SHIFT to the PRT and FT in the EMBARC study using a complete-case approach. Based on the EMBARC study design, baseline task completion was primarily driven by logistical factors independent of participants' task performance or behavioral responses; therefore, the missing completely at random assumption is likely to hold. Missingness also arose from factors unrelated to behavioral phenotypes, such as participants being unable to undergo neuroimaging because of contraindications (e.g., metal implants). Thus, although complete-case analysis has limitations, the study design suggests that missingness is unlikely to be systematically related to the behavioral parameters of interest. Additionally, the current analysis focused on the first block of the two tasks. It would be interesting in future work to develop methods that accommodate additional time-varying dynamics across blocks, enabling jointly analysis of all blocks.

SHIFT and corresponding identifiability discussion are developed for general $K$: the latent factor structure, the task-specific likelihoods, and the transition dynamics are all formulated without restriction to $K=2$. The use of two tasks in the empirical analysis is driven by the tasks available in the EMBARC study rather than by a limitation of the methodology. We agree, however, that additional practical considerations arise when $K>2$. For example, computational cost increases with $K$ because of repeated evaluation of task-specific likelihoods and latent-state processes. Due to the iterative nature of the EM algorithm and its reliance on multivariate Gauss–Hermite quadrature, SHIFT is computationally intensive and may not scale well when $K$, $N$, $J_{ik}$, $d_f$, and $R$ are large. Incorporating stochastic VA \citep{hoffman2013stochastic, hoffman2015structured} or black box VA \citep{ranganath2014black} may lead to more scalable and efficient implementations. Further, developing uncertainty quantification procedures for SHIFT would be of interest. Additionally, although the current approach uses binary latent states, a continuous representation of attentional states could provide a more nuanced characterization and represents an important direction for future work.

Another exciting opportunity for future research involves developing multimodal methods to integrate behavioral tasks with neuroimaging measures. By incorporating brain activity data, we could learn the neural mechanisms that underlie task engagement and focus, and how these relate to behavioral patterns, with the potential to improve the understanding of cognitive function and treatment responses for mental disorders. Our methodological framework also extends naturally to other applications such as joint modeling of multimodal sensor data, dynamic topic models, and latent variable settings where sequential dependence and shared structure must be captured simultaneously. 


\section*{Acknowledgments}
The authors thank the reviewers, the associate editor, and the co-editor for their valuable comments. This manuscript reflects the views of the authors and may not reflect the opinions or views of the National Institutes of Health or of the Submitters submitting original data to National Institute of Mental Health Data Archive.

\section*{Data Availability}
The EMBARC data can be obtained through an application at \url{https://dx.doi.org/10.15154/e99n-np66}.

\section*{Funding}
This research is supported by U.S. NIH grants NS073671, GM124104, and MH123487.

\bibliographystyle{apalike}
\bibliography{reference}

\end{document}